\documentclass[11pt]{article}
\usepackage{amsmath,amssymb,amsthm,amsxtra,overpic,bbm,bm,epsfig,ulem}
\usepackage{color}
\usepackage{multirow}
\usepackage{rotating}

\usepackage{array}
\newcommand{\PreserveBackslash}[1]{\let\temp=\\#1\let\\=\temp}
\newcolumntype{C}[1]{>{\PreserveBackslash\centering}p{#1}}
\newcolumntype{R}[1]{>{\PreserveBackslash\raggedleft}p{#1}}
\newcolumntype{L}[1]{>{\PreserveBackslash\raggedright}p{#1}}

\makeatletter

\newcommand{\Rmnum}[1]{\expandafter\@slowromancap\romannumeral #1@}
\makeatother

\begin{document}

\begin{center}
{\Large\bf A modified Friedberg-Lee symmetry for the TM1 neutrino mixing}
\end{center}

\vspace{0.1cm}

\begin{center}
{\bf Zhen-hua Zhao}\footnote{E-mail: zhaozhenhua@ihep.ac.cn}\\
{Institute of High Energy Physics, Chinese Academy of Sciences, Beijing 100049, China}
\end{center}

\vspace{1.5cm}

\begin{abstract}
In this letter, we put forward a special neutrino mass matrix which is invariant under a modified Friedberg-Lee (FL) transformation
$\nu^{}_e \to \nu^{}_e-2\xi$ and $\nu^{}_{\mu, \tau} \to \nu^{}_{\mu, \tau}+\xi$ with $\xi$
being a space-time independent element of the Grassmann algebra. Compared to the original FL symmetry (with the transformation
$\nu^{}_{e, \mu, \tau} \to \nu^{}_{e, \mu, \tau}+\xi$) which results in the TM2 neutrino mixing,
the modified FL symmetry will lead us to the TM1 mixing which has a better agreement with the experimental results.
While the original FL symmetry has to be broken in order to produce a realistic neutrino mass spectrum,
the modified FL symmetry is allowed to remain intact and give us a vanishing $m^{}_1$.
A combination of the FL symmetry with the $\mu$-$\tau$ reflection symmetry is also discussed.
\end{abstract}

\newpage

\framebox{1} \hspace{0.2cm}
Thanks to a variety of neutrino-oscillation experiments, the fact that (at least two) neutrinos are massive and lepton
flavors are mixed has been established \cite{PDG}. The lepton mixing arises from the mismatch between the flavor and mass eigenstates
of leptons. In the basis where the flavor eigenstates of charged leptons are identified with their mass eigenstates, the lepton mixing
is attributed to the unitary matrix $U$ needed for diagonalizing the neutrino mass matrix $M^{}_\nu$. Hereafter we will work in this
basis and refer to ``lepton mixing'' as ``neutrino mixing'' following the convention. In the standard picture of
three-neutrino mixing \cite{PMNS}, $U$ can be parameterized in the standard way as
\begin{equation}
U=\left( \begin{matrix} \vspace{0.1cm}
c_{12}c_{13} & s_{12}c_{13} & s_{13}e^{-i\delta} \cr \vspace{0.1cm}
-s_{12}c_{23}-c_{12}s_{23}s_{13}e^{i\delta} & c_{12}c_{23}-s_{12}s_{23}s_{13}e^{i\delta}  & s_{23}c_{13} \cr
s_{12}s_{23}-c_{12}c_{23}s_{13}e^{i\delta}  & -c_{12}s_{23}-s_{12}c_{23}s_{13}e^{i\delta} & c_{23}c_{13}
\end{matrix} \right)
\left( \begin{matrix}
e^{i\rho} & & \cr &e^{i\sigma} & \cr & & 1
\end{matrix} \right),
\label{1}
\end{equation}
where $s^{}_{ij}$ and $c^{}_{ij}$ stand for $\sin{\theta^{}_{ij}}$ and $\cos{\theta^{}_{ij}}$, $\delta$ is the Dirac phase while $\rho$ and
$\sigma$ are Majorana phases and only appear when neutrinos are of the Majorana type. Up to now, the three mixing angles as well as two
neutrino mass-squared differences have been measured to a good degree of accuracy. Different from the quark mixing, the neutrino mixing is
characteristic of two large ($\theta^{}_{23}$ is even possible to be maximal --- $\pi/4$) and one relatively small mixing angles.
For convenience, the global-fit results for these parameters are presented below \cite{GF}:
\begin{eqnarray}
&&\sin^2{\theta^{}_{12}}= 0.304^{+0.013}_{-0.012} \; , \hspace{0.3cm} \sin^2{\theta^{}_{23}}= 0.452^{+0.052}_{-0.028} \; , \hspace{0.3cm}
\sin^2{\theta^{}_{13}}= 0.0218^{+0.001}_{-0.001} \; ,  \nonumber \\
&&\Delta m^2_{31}=2.457^{+0.047}_{-0.047}\times 10^{-3} {\rm eV}^2 \; , \hspace{0.8cm}
\Delta m^2_{21}=7.50^{+0.19}_{-0.17}\times 10^{-5} {\rm eV}^2 \; .
\label{2}
\end{eqnarray}
In contrast, the result for $\delta$ is rather inconclusive and its allowed range remains (0, 2$\pi$) at the $3\sigma$ level.
Furthermore, there is not any information about the Majorana phases $\rho$ and $\sigma$.
Theoretically, how to interpret this salient neutrino mixing pattern
and predict the possible values of unknown parameters poses an interesting question.
Before the measurement of $\theta^{}_{13}$ \cite{DYB}, the Tri-Bimaximal (TBM) mixing \cite{TBM}
\begin{equation}
U^{}_{\rm TBM}= \left( \begin{matrix} \vspace{0.1cm}
\displaystyle\frac{\sqrt 2}{\sqrt 3} & \displaystyle\frac{1}{\sqrt 3} & 0 \cr \vspace{0.1cm}
-\displaystyle\frac{1}{\sqrt 6} & \displaystyle\frac{1}{\sqrt 3} & -\displaystyle\frac{1}{\sqrt 2} \cr
-\displaystyle\frac{1}{\sqrt 6} & \displaystyle\frac{1}{\sqrt 3} & \displaystyle\frac{1}{\sqrt 2}
\end{matrix} \right) \hspace{0.3cm} \Longrightarrow \hspace{0.3cm}
\left\{ \begin{array}{l} \vspace{0.15cm}
\sin^2{\theta^{}_{12}}=\displaystyle\frac{1}{3} \; ,\\ \vspace{0.15cm}
\sin^2{\theta^{}_{23}}=\displaystyle\frac{1}{2} \; , \\ \sin^2{\theta^{}_{13}}=0 \;,
\end{array} \right.
\label{3}
\end{equation}
which agreed well with the observations was widely accepted as the correct description of neutrino mixings.
Due to the simplicity and prediction power of the TBM mixing, many people believe that some flavor symmetries must exist
in the lepton sector. In this respect, numerous flavor symmetries such as $A^{}_4$ and $S^{}_4$ \cite{FS} have been
employed to construct realistic models to derive the TBM neutrino mixing. However, the discovery of a relatively large $\theta^{}_{13}$
rejects this simple situation. As possible alternatives to the TBM mixing, the TM1 and TM2 mixing patterns have received a lot
of attention \cite{TM12}:
\begin{equation}
U^{}_{\rm TM1}=U^{}_{\rm TBM} \left( \begin{matrix}
1 & &  \cr & c & s e^{-i \phi} \cr & -s e^{i\phi} & c
\end{matrix} \right), \hspace{0.5cm}
U^{}_{\rm TM2}=U^{}_{\rm TBM} \left( \begin{matrix}
c & & s e^{-i \phi} \cr & 1 &  \cr -s e^{i\phi} &  & c
\end{matrix} \right),
\label{4}
\end{equation}
with $c=\cos{\theta}$ and $s=\sin{\theta}$.
They are phenomenological viable in the sense that a non-zero $\theta^{}_{13}$ is generated by $\theta$ while $\theta^{}_{12}$
stays close to its value in the TBM case and $\theta^{}_{23}$ can remain nearly maximal when $\phi$ approaches $\pi/2$ or $3\pi/2$.
It is therefore desirable for us to explore appropriate physical scenarios (especially flavor symmetries) to justify these two specific
mixing patterns \cite{TM2,TM1}. Interestingly, the Friedberg-Lee (FL) symmetry \cite{FL} proposed long before happens to result in the TM2 mixing.
In this letter, we aim to modify the original FL symmetry so as to accommodate the TM1 mixing. This is partly motivated by the fact
that the value of $\theta^{}_{12}$ in the TM1 case is much closer to the observed value than that in the TM2 case.
As we will see, the modified FL symmetry also has an advantage in fitting with the neutrino mass spectrum over the original one.

\framebox{2} \hspace{0.2cm}
First of all, let us briefly recapitulate the main characteristics and consequences of
the FL symmetry applied to the neutrino sector \cite{FL,XZZ,FL2} \footnote{For its application to the quark sector, see \cite{Quark}. }.
Assuming neutrinos to be Majorana fermions, their mass operators are enforced the following form by the FL symmetry \cite{FL}
\begin{eqnarray}\label{}
\mathcal{L}^{}_{\rm mass} &= & a (\eta^{*}_\tau \bar \nu^{}_\mu  - \eta^{*}_\mu \bar \nu^{}_\tau )
(\eta^{*}_\tau \nu^{c}_\mu - \eta^{*}_\mu \nu^{c}_\tau)+ b(\eta^{*}_\mu \bar \nu^{}_e -\eta^{*}_e \bar \nu^{}_\mu )
(\eta^{*}_\mu \nu^{c}_e -\eta^{*}_e \nu^{c}_\mu ) + \nonumber \\
& & d (\eta^{*}_\tau \bar \nu^{}_e -\eta^{*}_e \bar \nu^{}_\tau ) (\eta^{*}_\tau \nu^{c}_e - \eta^{*}_e \nu^{c}_\tau)+ {\rm h.c.} \; ,
\label{5}
\end{eqnarray}
with $a$, $b$, $d$ and $\eta^{}_{e,\mu,\tau}$ being complex numbers.
It is easy to check that $\mathcal{L}^{}_{\rm mass}$ keeps unchanged under the FL symmetry transformation \cite{FL}
\begin{equation}
\nu^{}_e \to \nu^{}_e +\eta^{}_e \xi \; , \hspace{0.5cm} \nu^{}_\mu \to \nu^{}_\mu +\eta^{}_\mu \xi \; ,
\hspace{0.5cm} \nu^{}_\tau \to \nu^{}_\tau +\eta^{}_\tau \xi \; ,
\label{6}
\end{equation}
where $\xi$ is a space-time independent element of the Grassmann algebra and anti-commutates with the neutrino fields.
When the FL symmetry was originally proposed, Friedberg and Lee just focused on the simplest case
$\eta^{}_e=\eta^{}_\mu=\eta^{}_\tau$ which results in a neutrino mass matrix as \cite{XZZ}
\begin{equation}\label{}
M^{}_{\rm FL}=\left( \begin{matrix}
b+d & -b & -d \cr -b & a+b & -a \cr -d & -a & a+d
\end{matrix} \right).
\label{7}
\end{equation}
Noteworthy, it can be transformed into a simple form by the TBM matrix
\begin{equation}\label{}
U^{\dagger}_{\rm TBM} M^{}_{\rm FL} U^{*}_{\rm TBM}=\left( \begin{matrix}
\displaystyle \frac{3}{2}(b+d) & \hspace{0.2cm}0 & \hspace{0.2cm} \displaystyle \frac{\sqrt{3}}{2}(b-d) \cr
0 & \hspace{0.2cm} 0 & \hspace{0.2cm} 0 \cr
\displaystyle \frac{\sqrt{3}}{2}(b-d) & \hspace{0.2cm} 0 & \hspace{0.2cm} 2a+\displaystyle \frac{1}{2}(b+d)
\end{matrix} \right).
\label{8}
\end{equation}
Apparently, the unitary matrix for diagonalizing $M^{}_{\rm FL}$ will be simply the TBM if one has $b=d$.
Otherwise, it will take the TM2 form as that given in Eq. (\ref{4})
\begin{equation}
U^{}_{\rm TM2}=\left( \begin{matrix} \vspace{0.1cm}
\displaystyle\frac{\sqrt 2}{\sqrt 3} c & \hspace{0.2cm}\displaystyle\frac{1}{\sqrt 3}
& \hspace{0.2cm} \displaystyle\frac{\sqrt 2}{\sqrt 3} s e^{-i \phi} \cr \vspace{0.1cm}
-\displaystyle\frac{1}{\sqrt 6} c + \displaystyle \frac{1}{\sqrt{2}}s e^{i \phi} & \hspace{0.2cm} \displaystyle\frac{1}{\sqrt 3}
& \hspace{0.2cm} -\displaystyle\frac{1}{\sqrt 6}s e^{-i \phi} -\displaystyle\frac{1}{\sqrt 2} c \cr
-\displaystyle\frac{1}{\sqrt 6} c - \displaystyle \frac{1}{\sqrt{2}}s e^{i \phi} & \hspace{0.2cm} \displaystyle\frac{1}{\sqrt 3}
& \hspace{0.2cm} -\displaystyle\frac{1}{\sqrt 6}s e^{-i \phi} + \displaystyle\frac{1}{\sqrt 2} c
\end{matrix} \right)
\left( \begin{matrix}
e^{i\varphi^{}_1} & & \cr & 1 & \cr & & e^{i \varphi^{}_3}
\end{matrix} \right),
\label{9}
\end{equation}
where $\varphi^{}_{1,3}$ are the Majorana phases needed to make the mass eigenvalues real and positive.
Taking into account the fact that $s$ should be equal to $\sqrt{3}s^{}_{13}/\sqrt{2}$, we obtain the $\theta^{}_{12}$ as
\begin{eqnarray}
s^2_{12}= \frac{1}{3}\frac{1}{1-s^2_{13}} =0.341 \;,
\label{10}
\end{eqnarray}
which is considerably larger than the observed value given in Eq. (\ref{2}). As mentioned, this is one of the reasons why we want to go beyond
the original FL symmetry (namely the case with $\eta^{}_e=\eta^{}_\mu=\eta^{}_\tau$). The other one is that the neutrino mass matrix
given by Eq. (\ref{7}) fails to give a realistic neutrino mass spectrum. As one can see from Eq. (\ref{8}), the resultant $m^{}_2$ would be vanishing
in conflict with the well-established fact $m^2_2-m^2_1>0$. Consequently, the FL symmetry has to be broken in order to produce a non-zero $m^{}_2$.
The commonly used way is to add a universal mass term which explicitly breaks the FL symmetry \cite{FL} to the Lagrangian relevant for
neutrino masses
\begin{equation}\label{}
\mathcal{L}^{}_{\rm mass} \to \mathcal{L}^{}_{\rm mass} + m^{}_0(\bar \nu^{}_e \nu^{c}_e +
\bar \nu^{}_\mu \nu^{c}_\mu + \bar \nu^{}_\tau \nu^{c}_\tau) + {\rm h.c.} \; .
\label{11}
\end{equation}
A detailed study about the corresponding consequences has been presented in \cite{XZZ},
so we shall directly turn to the case with a modified FL symmetry.

\framebox{3} \hspace{0.2cm}
The main purpose of this letter is to show that a modified FL symmetry with $\eta^{}_e=-2\eta^{}_\mu=-2\eta^{}_\tau$ can address
the problems encountered in the case with $\eta^{}_e=\eta^{}_\mu=\eta^{}_\tau$. In this case, the neutrino mass matrix turns out to be
\begin{equation}\label{}
M^\prime_{\rm FL}=\left( \begin{matrix}
b+d & 2b & 2d \cr 2b & a+4b & -a \cr 2d & -a & a+4d
\end{matrix} \right),
\label{12}
\end{equation}
which can be transformed into the following form by the TBM matrix
\begin{equation}\label{}
U^{\dagger}_{\rm TBM} M^\prime_{\rm FL} U^{*}_{\rm TBM}=\left( \begin{matrix}
0 & \hspace{0.2cm}0 & \hspace{0.2cm} 0 \cr
0 & \hspace{0.2cm}3(b+d) & \hspace{0.2cm} \sqrt{6}(d-b) \cr
0 & \hspace{0.2cm}\sqrt{6}(d-b) & \hspace{0.2cm} 2(a+b+d)
\end{matrix} \right).
\label{13}
\end{equation}
Similarly, the unitary matrix for the diagonalization of $M^\prime_{\rm FL}$ will also be the TBM if we have $b=d$.
In the case of $b \neq d$, $M^\prime_{\rm FL}$ will be diagonalized by a unitary matrix of the form
\begin{equation}
U^{}_{\rm TM1}=\left( \begin{matrix} \vspace{0.1cm}
\displaystyle\frac{\sqrt 2}{\sqrt 3}  & \hspace{0.2cm}\displaystyle\frac{1}{\sqrt 3} c
& \hspace{0.2cm} \displaystyle\frac{1}{\sqrt 3} s e^{-i \phi} \cr \vspace{0.1cm}
-\displaystyle\frac{1}{\sqrt 6}   & \hspace{0.2cm} \displaystyle\frac{1}{\sqrt 3}c+ \displaystyle \frac{1}{\sqrt2}s e^{i\phi}
& \hspace{0.2cm} \displaystyle\frac{1}{\sqrt 3}s e^{-i \phi} -\displaystyle\frac{1}{\sqrt 2} c \cr
-\displaystyle\frac{1}{\sqrt 6}  & \hspace{0.2cm} \displaystyle\frac{1}{\sqrt 3}c-\displaystyle \frac{1}{\sqrt 2} s e^{i\phi}
& \hspace{0.2cm} \displaystyle\frac{1}{\sqrt 3}s e^{-i \phi} + \displaystyle\frac{1}{\sqrt 2} c
\end{matrix} \right)
\left( \begin{matrix}
1 & & \cr & e^{i \varphi^{}_2} & \cr & & e^{i \varphi^{}_3}
\end{matrix} \right),
\label{14}
\end{equation}
where $\theta$, $\phi$ and $\varphi^{}_{2,3}$ are determined by
\begin{eqnarray}
&&\tan{2\theta} = \frac{\sqrt{X^2+Y^2}}{Z} \;,  \hspace{1cm}\tan{\phi} = \frac{X}{Y} \;, \nonumber \\
&&\varphi^{}_{2}= -\frac{1}{2} {\rm Arg}\left[3(b+d)c^2-2 \sqrt{6}(d-b)c s e^{-i\phi} +2(a+b+d) s^2 e^{-2i\phi} \right] \;, \nonumber \\
&&\varphi^{}_3 = -\frac{1}{2} {\rm Arg}\left[ 3(b+d)s^2 e^{2i\phi}+2\sqrt{6}(d-b)c s e^{i\phi} +2(a+b+d) c^2 \right] \;,
\label{15}
\end{eqnarray}
with
\begin{eqnarray}
&& X = 2\sqrt{6} {\rm Im}[(2a-b-d)(d-b)^*] \;, \hspace{1cm} Y = 2\sqrt{6}{\rm Re}[(2a+5b+5d)(d-b)^*]\;, \nonumber \\
&& Z ={\rm Re}[(2a-b-d)(2a+5b+5d)^*] \;.
\label{16}
\end{eqnarray}
In addition, the neutrino masses are given by
\begin{eqnarray}
m^{}_2 &=& \left|3(b+d)c^2-2 \sqrt{6}(d-b)c s e^{-i\phi} +2(a+b+d) s^2 e^{-2i\phi}\right|  \;, \nonumber \\
m^{}_3 &=& \left|3(b+d)s^2 e^{2i\phi}+2\sqrt{6}(d-b)c s e^{i\phi} +2(a+b+d) c^2 \right| \;,
\label{17}
\end{eqnarray}
with $m^{}_1=0$ as a consequence of the FL symmetry. Because of the vanishing of $m^{}_1$, only the difference between $\varphi^{}_2$ and
$\varphi^{}_3$ is of physical meaning. By comparing Eq. (\ref{14}) with Eq. (\ref{1}),
one obtains the standard-parametrization mixing angles as
\begin{eqnarray}
&&s^{2}_{13}= \frac{1}{3}s^2 \;, \hspace{1cm} s^2_{12}=\frac{1}{3}\frac{1-3s^2_{13}}{1-s^2_{13}}=0.318 \;, \nonumber \\
&&s^2_{23} = \frac{1-s^2_{13}-2\sqrt{2(1-3s^2_{13})}s^{}_{13}\cos{\phi}}{2(1-s^2_{13})}\approx \frac{1}{2}(1-2\sqrt{2}s^{}_{13}\cos{\phi}) \;.
\label{18}
\end{eqnarray}
Without saying, $s$ is required to be $\sqrt{3}s^{}_{13}=0.256$ for Eq. (\ref{14}) to be phenomenological viable. Most importantly,
$s^2_{12}$ turns out to have a value of 0.318 which as promised is in better agreement with the measurements than that obtained
in the original FL symmetry case. On the other hand, $s^{2}_{23}$ is not completely fixed but dependent on $\phi$ as
illustrated in Fig. 1(a). For comparison, the $1\sigma$ range of $s^{2}_{23}$ is also shown. One can see that
$\phi$ is constrained to be around $\pi/2$ or $3\pi/2$, specifically in the ranges (1.2, 1.6) or (4.7, 5.1).
Furthermore, the Jarlskog invariant which measures the strength of CP violation in neutrino oscillations is found to be
\begin{eqnarray}
J=-\frac{1}{3\sqrt{2}}\sqrt{1-3s^2_{13}}s^{}_{13}\sin{\phi} \;,
\label{19}
\end{eqnarray}
which has a value in the ranges $\pm$(0.0314, 0.0336) after the constraint on $\phi$ has been taken into consideration.
By comparing the expression of $J$ in Eq. (\ref{19}) with that in terms of the standard-parametrization parameters
(i.e., $J=c^{}_{12}s^{}_{12} c^2_{13}s^{}_{13}c^{}_{23}s^{}_{23}\sin{\delta}$), we get the Dirac phase $\delta$ as
\begin{eqnarray}
\frac{\sin{\delta}}{\sin{\phi}}=\frac{1-s^2_{13}}{\sqrt{4s^4_{13}+(1-3s^2_{13})^2-4(1-3s^2_{13})s^2_{13}\cos{2\phi}}} \;.
\label{20}
\end{eqnarray}
The right-hand side of this equation ranges from 1 to 1.1 as indicated by Fig. 1(b). In particular, $\delta$ will be
exactly equal to $\phi$ for $\phi=\pi/2$ or $3\pi/2$ from which the constrained $\phi$ is not far. Hence $\delta$ can be
identified with $\phi$ as a good approximation. As an immediate consequence, Eq. (\ref{18}) can be viewed as a correlation
between $\theta^{}_{23}$ and $\delta$ which will be tested in the near future. As for the phases $\varphi^{}_{2,3}$, no
definite conclusions can be drawn for them.

\begin{figure}
\includegraphics[width=\textwidth]{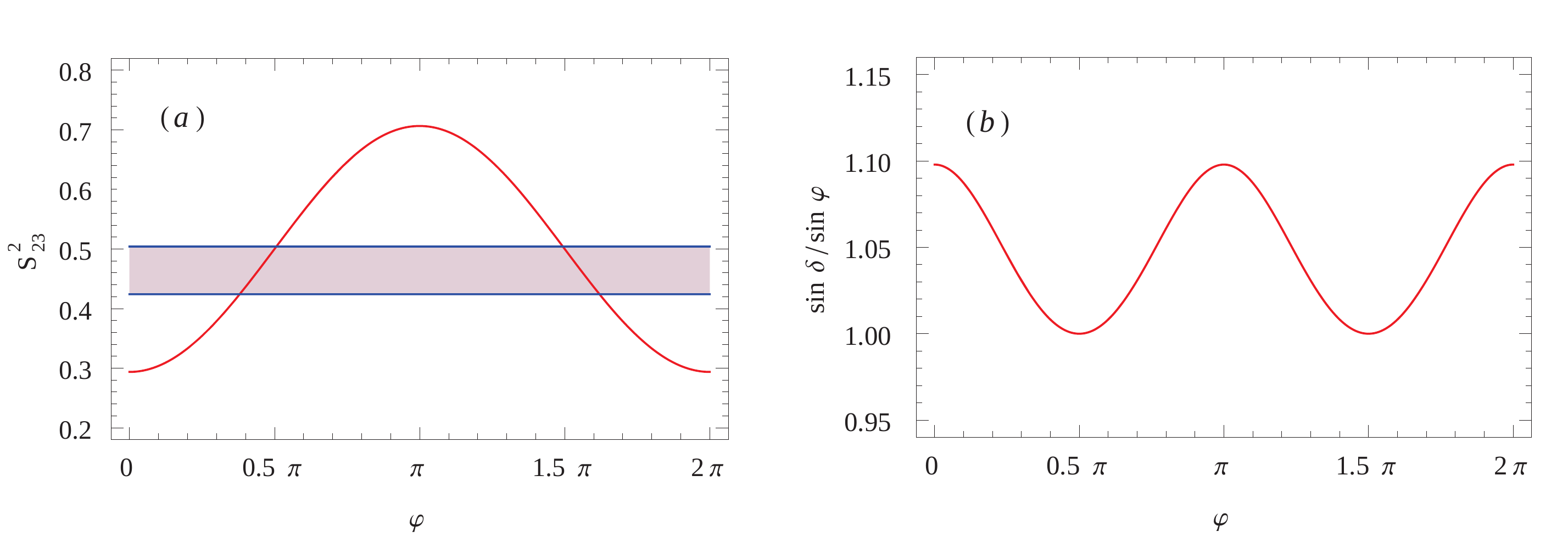}
\caption{The possible values of $s^{2}_{23}$ and $\sin{\delta}/\sin{\phi}$ against $\phi$. The horizontal band in the left figure
shows the $1\sigma$ range of $s^{2}_{23}$ obtained in the global fit.}
\end{figure}

\framebox{4} \hspace{0.2cm}
Finally, we point out that a combination of the FL symmetry with the $\mu$-$\tau$ reflection symmetry will bring about more definite results
(including those relevant for $\varphi^{}_{2,3}$). The $\mu$-$\tau$ reflection symmetry \cite{MTR,MTR2} is defined in the sense
that $M^{}_\nu$ should be invariant under the transformation
\begin{eqnarray}
\nu^{}_{e} \rightleftharpoons \nu^{c}_{e }\ , \hspace{0.5cm} \nu^{}_{\mu } \rightleftharpoons
\nu^{c}_{\tau}\ , \hspace{0.5cm} \nu^{}_{\tau } \rightleftharpoons \nu^{c}_{\mu} \;,
\label{21}
\end{eqnarray}
which imposes the following conditions on the elements of $M^{}_\nu$
\begin{eqnarray}
M^{}_{e\mu}=M^{*}_{e\tau} \;, \hspace{0.5cm} M^{}_{\mu\mu}=M^{*}_{\tau\tau} \;,
\hspace{0.5cm} {\rm Im}(M^{}_{ee})={\rm Im}(M^{}_{\mu\tau})=0 \;.
\label{22}
\end{eqnarray}
This symmetry is a kind of generalized CP symmetry \cite{GCP} which combines the charge conjugation and $\mu$-$\tau$ interchange operations.
When it is combined with the FL symmetry in the case $\eta^{}_e=-2\eta^{}_\mu=-2\eta^{}_\tau$,
the neutrino mass matrix is restricted to the form
\begin{equation}\label{}
M^{\prime\prime}_{\rm FL}=\left( \begin{matrix}
b+b^* & 2b & 2b^* \cr 2b & a+4b & -a \cr 2b^* & -a & a+4b^*
\end{matrix} \right),
\label{23}
\end{equation}
where the parameter $a$ is real. Note that it is the equality of $\eta^{}_\mu$ and $\eta^{}_\tau$ that allows this combination.
Otherwise, a combination of the FL symmetry with the $\mu$-$\tau$ reflection symmetry will not make sense.
The corresponding neutrino mixing matrix still takes the form as parameterized in Eq. (\ref{14}) but with $\theta$ and $\phi$
given by
\begin{eqnarray}
\tan{2\theta} e^{i\phi} =i\frac{2\sqrt{6}{\rm Im}(b)}{a+{\rm Re}(b)} \;,
\label{24}
\end{eqnarray}
while $\varphi^{}_{2,3}$ should take appropriate values that make the mass eigenvalues positive:
\begin{eqnarray}
m^{}_2  &=& \left[6{\rm Re}(b)c^2 \pm 4 \sqrt{6}{\rm Im}(b) c s -2(a+2{\rm Re}(b)) s^2 \right]e^{2i\varphi^{}_2} \;, \nonumber \\
m^{}_3  &=& \left[-6{\rm Re}(b)s^2 \pm 4\sqrt{6}{\rm Im}(b)c s +  2(a+2{\rm Re}(b)) c^2 \right]e^{2i\varphi^{}_3} \;.
\label{25}
\end{eqnarray}
The sign difference in $m^{}_{2,3}$ arises from the two possible choices of $\phi$ --- $\pi/2$ and $3\pi/2$.
One can easily draw the conclusion that $\phi$ and equivalently $\delta$ have a value of $\pi/2$ or $3\pi/2$ (in which cases
$\theta^{}_{23}$ is fixed to $\pi/4$) and $\varphi^{}_{2,3}$ are simply 0 or $\pi/2$. Once the choices for $\phi$ and $\varphi^{}_{2,3}$
are specified, all the three free parameters $a$, ${\rm Re}(b)$ and ${\rm Im}(b)$ can be numerically determined.
For example, the combination ($\pi/2$, 0, 0) of ($\phi$, $\varphi^{}_2$, $\varphi^{}_3$) is a consequence of
\begin{eqnarray}
a=0.0214 {\rm eV}\;, \hspace{1cm} {\rm Re}(b)=0.0010{\rm eV} \;, \hspace{1cm} {\rm Im}(b)=0.0026{\rm eV}\;.
\end{eqnarray}
In short, we can pin down all the mixing parameters as well as neutrino masses in the scenario
where the FL symmetry coexists with the $\mu$-$\tau$ reflection symmetry.

\framebox{5} \hspace{0.2cm}
In conclusion, we have proposed a new application of the Friedberg-Lee symmetry in the neutrino sector to replace the original one.
Although neutrino fields were originally assumed to transform universally under the FL symmetry (i.e., $\nu^{}_{e, \mu, \tau}
\to \nu^{}_{e, \mu, \tau} +\xi$) \cite{FL}, we advocate that they may transform in the non-universal way $\nu^{}_{e} \to \nu^{}_e -2 \xi$
and $\nu^{}_{\mu, \tau} \to  \nu^{}_{\mu, \tau}+ \xi$ for phenomenological considerations. By means of such a slight modification,
the phenomenological consequences are improved in two aspects: (1) The original FL symmetry would lead to a neutrino mass spectrum with $m^{}_2=0$
which contradicts the fact $m^{2}_2-m^2_1>0$ and thus needs to be broken. In comparison, the modified FL symmetry does not necessarily break
since it gives us a vanishing $m^{}_1$ which is still an allowed possibility. (2) As for neutrino mixings, the original TM2 mixing is replaced
by the TM1 one which is more favored by the observations. To be specific, $s^{2}_{12}$ is predicted to have a value of 0.341 in the TM2 case
but 0.318 in the TM1 case while its measured value is in the ranges 0.292-0.317 and 0.270-0.344 at the $1\sigma$ and $3\sigma$ levels respectively.
Moreover, a correlation between $\theta^{}_{23}$ and $\delta$ exists, and it constrains $\delta$ to lie around $\pi/2$ (1.2-1.6)
or $3\pi/2$ (4.7-5.1). Precise measurements for $\theta^{}_{23}$ and $\delta$ are demanding to confirm or rule out this correlation.
In addition, we have also discussed the possibility of combing the FL and $\mu$-$\tau$ reflection symmetries so that
all the physical parameters related to neutrinos can be fixed. So far all the predictions of the modified FL symmetry are consistent with
the current data, so we will not bother to discuss its breaking. In the next step, we should study the possible origin of the FL symmetry
which is still unclear to us at the moment.

\vspace{1cm}

\underline{Acknowledgments} \hspace{0.2cm} I am very grateful to Professor Zhi-zhong Xing for bringing the Friedberg-Lee symmetry into
my attention, encouraging me to carry out such a study and reading the manuscript. This work was supported in part by the National Natural Science
Foundation of China under Nos. 11375207 and 11135009 and by the China Postdoctoral Science Foundation under
Grant No. 2015M570150.

\end{document}